\documentclass[aps,prl,twocolumn,nobibnotes,superscriptaddress,notitlepage]{revtex4-1}
\usepackage[hyperindex,breaklinks,colorlinks,urlcolor=purple,citecolor=purple,linkcolor=purple]{hyperref}
\usepackage{graphicx}
\usepackage{natbib}
\usepackage{subfigure}
\usepackage{amsmath}
\usepackage{dcolumn}
\usepackage{bm}
\usepackage[dvipsnames]{xcolor}

\newcommand{\huPhys}{Department of Physics, Harvard University, Cambridge, Massachusetts 02138, USA}
\newcommand{\huChem}{Department of Chemistry and Chemical Biology, Harvard University, Cambridge, Massachusetts 02138, USA}
\newcommand{\umdPhys}{Department of Physics, University of Maryland, College Park, Maryland 20742, USA}
\newcommand{\umdEE}{Department of Electrical and Computer Engineering, University of Maryland, College Park, Maryland 20742, USA}
\newcommand{\umdQTC}{Quantum Technology Center, University of Maryland, College Park, Maryland 20742, USA}
\newcommand{\cfa}{Harvard-Smithsonian Center for Astrophysics, Cambridge,	Massachusetts	02138,	USA}
\newcommand{\ncsuchem}{Department of Chemistry, North Carolina State University,	Raleigh, North Carolina	27695,	USA}
\newcommand{\ncsuphy}{Department of Physics, North Carolina State University,	Raleigh, North Carolina	27695,	USA}
\newcommand{\unc}{UNC $\&$ NC State Joint Department of Biomedical Engineering, Raleigh, North Carolina	27695,	USA}
\newcommand{\cbs}{Center for Brain Science, Harvard	University,	Cambridge,	Massachusetts	02138,	USA}
\newcommand{\AAM}{A. A. Martinos Center for Biomedical Imaging, Massachusetts General Hospital, Boston, Massachusetts 02129, USA}
\newcommand{\HMS}{Harvard Medical School, Boston, Massachusetts 02129, USA}
\newcommand{\tum}{Department of Chemistry, Technical University of Munich, Germany}

\begin{document}

\title{Micron-scale SABRE-enhanced NV-NMR Spectroscopy}
\date{\today}

\author{Nithya Arunkumar}
\affiliation{\huPhys}
\affiliation{\umdQTC}
\affiliation{\cfa}

\author{Dominik B. Bucher}
\affiliation{\huPhys}
\affiliation{\cfa}
\affiliation{\tum}

\author{Matthew J. Turner} 
\affiliation{\huPhys}
\affiliation{\cbs}

\author{Patrick TomHon}
\affiliation{\ncsuchem}

\author{David Glenn}
\affiliation{\huPhys}

\author{Soren Lehmkuhl}
\affiliation{\ncsuchem}

\author{Mikhail D. Lukin}
\affiliation{\huPhys}

\author{Hongkun Park}
\affiliation{\huPhys}
\affiliation{\huChem}

\author{Matthew S. Rosen}
\affiliation{\huPhys}
\affiliation{\AAM}
\affiliation{\HMS}

\author{Thomas Theis}
\affiliation{\ncsuchem}
\affiliation{\ncsuphy}
\affiliation{\unc}

\author{Ronald L. Walsworth}
\thanks{Correspondence to: walsworth@umd.edu}
\affiliation{\huPhys}
\affiliation{\umdQTC}
\affiliation{\cfa}
\affiliation{\cbs}
\affiliation{\umdPhys}
\affiliation{\umdEE}

\date{\today}

\begin{abstract}

Optically-probed nitrogen-vacancy (NV) quantum defects in diamond can detect nuclear magnetic resonance (NMR) signals with high-spectral resolution from micron-scale sample volumes of about 10 picoliters. However, a key challenge for NV-NMR is detecting samples at millimolar concentrations. Here, we demonstrate an improvement in NV-NMR proton concentration sensitivity of about $10^5$ over thermal polarization by hyperpolarizing sample proton spins through signal amplification by reversible exchange (SABRE), enabling micron-scale NMR of small molecule sample concentrations as low as 1 millimolar in picoliter volumes. The SABRE-enhanced NV-NMR technique may enable detection and chemical analysis of low concentration molecules and their dynamics in complex micron-scale systems such as single-cells.	

\end{abstract}

\maketitle

Nitrogen vacancy (NV) quantum defects in diamond are a leading modality for sensitive magnetometry with high spatial-resolution and operation under ambient conditions~\cite{Taylor2008,Barry2020}, including for nuclear magnetic resonance (NMR) spectroscopy at small length scales (nanometers to microns)~\cite{Staudacher2013, Mamin2013,Muller2014, Sushkov2014, Lovchinsky2016, Aslam2017}. Initial work on NV-NMR spectroscopy~\cite{Staudacher2013, Mamin2013,Muller2014,Sushkov2014,Lovchinsky2016,Aslam2017} suffered from low spectral resolution (kHz), due to the short decoherence time of the NV centers. To overcome this problem, Glenn \textit{et al.}~\cite{Glenn2018} implemented a coherently averaged synchronized readout (CASR) technique and demonstrated an NV-NMR spectral resolution of a few Hz at 88 mT on a micron-scale sensing volume. However, due to the finite sensitivity of the NV-NMR sensor, its application is restricted to highly concentrated pure samples, which limits its utility for most chemical and biological problems. 

Recently, Bucher \textit{et al.}~\cite{Bucher2018} employed dynamic nuclear polarization (DNP) based on the Overhauser mechanism~\cite{Overhauser1953}, where polarization is transferred to the sample nuclear spins from the electronic spins of dissolved molecular radicals, and obtained two orders of magnitude proton number sensitivity enhancement for micron-scale CASR NV-NMR. However, the DNP sensitivity enhancement is limited by the finite electronic spin polarization at the low magnetic fields and ambient temperatures used for NV-NMR. Higher sample nuclear spin polarization and thereby improvement in NMR sensitivity can be achieved through a parahydrogen-based signal amplification by reversible exchange (SABRE) process~\cite{Adams2009,Cowley2011,RaynerE2017}. To date SABRE has been shown to enhance conventional inductive NMR sensitivity in mL to $\mu$L scale sensing volumes~\cite{Adams2009,Cowley2011,RaynerE2017,Theis2012,Gong2010,Theis2015}.

\begin{figure*}[ht]
	\centering
	\includegraphics[width=7 in]{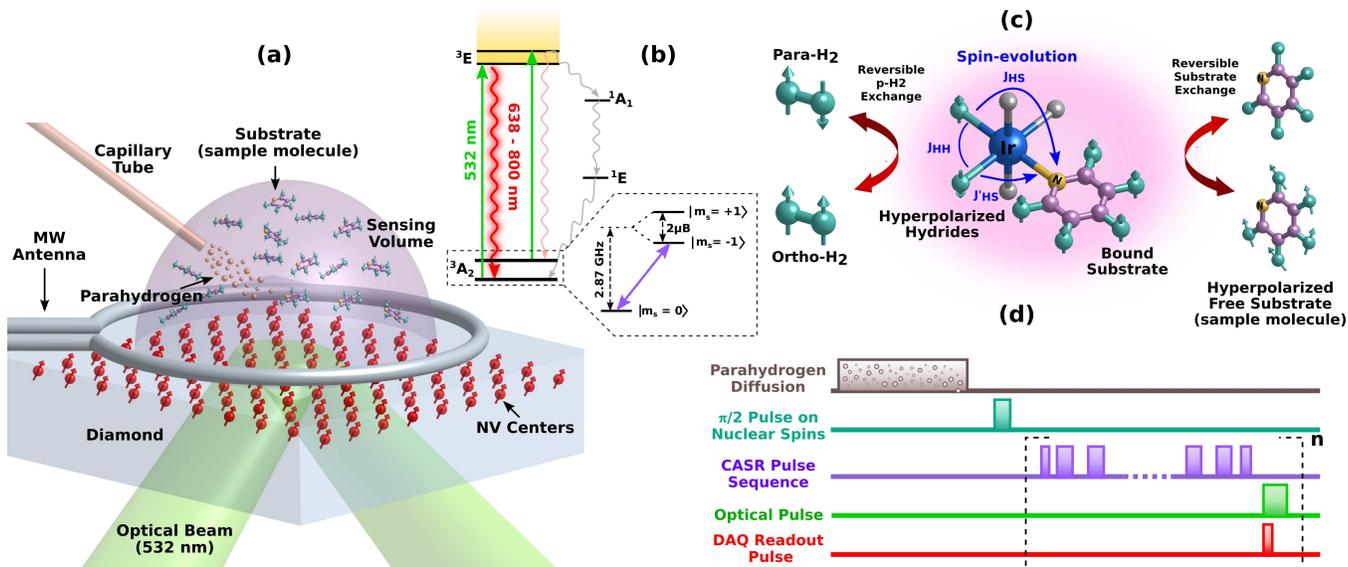} 
	\caption{
		\textbf{NV-NMR sensor integrated with signal amplification by reversible exchange (SABRE). (a)} Experimental schematic of the ensemble NV-NMR sensor. A 532 nm optical beam illuminates the diamond chip via a total internal reflection configuration. A microwave antenna on the diamond chip drives the electronic spins of the NV centers. Parahydrogen diffused into the sample through the capillary tube, initiates the SABRE reaction, and thereby hyperpolarizes sample proton spins via an Iridium-based catalyst in the sample solution. The hyperpolarized NMR signal is sensed using the electronic spins of the ensemble of NV centers. \textbf{(b)} Energy level diagram for the nitrogen vacancy (NV) centers in diamond. The expanded view shows the Zeeman splitting of the ground triplet state $^3$A$_2$ in the presence of a magnetic field.   \textbf{(c)} SABRE hyperpolarization process. The SABRE catalyst is in reversible exchange (indicated by red arrows) with parahydrogen (left side) and a small molecule substrate (e.g., pyridine, right side). In the transient (ms) bound state of the catalyst-substrate complex (center), spin order flows from the hydrides to the substrate leading to polarization built up on the free substrate in solution, which is the sample to be probed with NV-NMR. \textbf{(d)} Pulse sequence for SABRE hyperpolarization and NV-NMR detection. The parahydrogen, bubbled into the sample solution, activates the catalyst and hyperpolarizes the small molecule substrate (the sample). A $\pi$/2 pulse induces a free nuclear precession (FNP) signal from the hyperpolarized sample, which is detected by NV sensor spins via a coherently-averaged synchronized readout (CASR) pulse sequence. }
	\label{fig:fig1}
\end{figure*}

Here, we integrate SABRE hyperpolarization with CASR NV-NMR to realize five orders of magnitude enhancement in concentration sensitivity relative to thermal nuclear spin polarization - in a micron-scale sample. Using our method, we measure the NMR spectrum from small molecule samples with concentrations as low as 1 millimolar and at a sensing volume of 10 picoliters. The high spectral resolution of SABRE NV-NMR enables the measurement of \textit{J}-couplings in dilute molecules, thereby providing chemical specificity at a low magnetic field of about 6.6 mT.

The experimental set up is shown in Fig.~\ref{fig:fig1}(a). The NV-NMR sensor is a ($2\times2\times0.5$)$\,$mm$^3$ high purity diamond chip with 13 $\mu$m thick NV layer and an NV concentration of $3\times10^{17} $cm$^{-3}$. The [111] axis of the NV-NMR sensor is oriented parallel to the bias magnetic field ($B_0$ $\approx$ 6.6 mT), which is generated by a feedback-stabilized electromagnet~\cite{Glenn2018}. A green optical beam ($\lambda$ = 532 nm) in a total internal reflection configuration with a spot diameter of $\sim$15 $\mu$m is used to initialize and readout the electronic spins of the NV-NMR sensor~\cite{Glenn2018}. A single-coil wire loop antenna~\cite{Bucher2019}, placed directly above the NV surface of the diamond is used to drive the electron spin resonance transitions of the NV centers. At the proton NMR frequency of 280 kHz for a 6.576 mT bias field, the AC magnetic field sensitivity ($\eta_B$) of the NV ensemble sensor is 35(2) pT/$\sqrt{Hz}$~\cite{SI}. The liquid sample is placed directly on top of the diamond surface. The thickness of the NV layer and the spot diameter of the optical beam provides an effective NMR sensing volume of about 10 pL~\cite{Glenn2018,Bucher2018}. 

Hyperpolarization of proton spins in the sample molecules is obtained through SABRE. Parahydrogen gas is first dispersed into the sample solution (Fig.~\ref{fig:fig1}(a)) for about 20 minutes to activate an Iridium-based catalyst~\cite{SI}, which then mediates reversible exchange of spin order between the parahydrogen and the small molecule substrate - the sample to be probed with NV-NMR - as shown in Fig.~\ref{fig:fig1}(c). Once activated, about 30 s of additional parahydrogen bubbling is sufficient to establish hyperpolarization on the substrate. During the transient lifetime of the catalyst-substrate complex (on the order of ms), proton spin order flows from the hydrides (in parahydrogen) to protons in the bound small molecule substrate (e.g., pyridine). Lastly, the hyperpolarized substrate dissociates, to give free hyperpolarized small molecules in solution with polarization lifetime T$_1$ $\sim$ 5 s. The polarization transfer process is resonant at about 6.6 mT, where the \textit{J}-coupling between the hydrides equals the frequency difference between hydride and substrate proton spins, leading to a level-anti-crossing between the singlet state of the hydrides and the proton spin-down states of the substrate. In summary, spin evolution and chemical exchange continually pump hyperpolarization into free small molecules in solution, as long as the parahydrogen is periodically refreshed by bubbling between NV-NMR measurements. Details of the home-built parahydrogen generation, its integration with the NV-NMR sensor, and the SABRE polarization transfer mechanism are discussed in the supplementary material~\cite{SI}.

After a one second wait time following SABRE hyperpolarization, a $\pi /2$ RF pulse is applied resonant with the nuclear spins of the sample. The induced Larmor precession of the nuclear spin results in a decaying oscillatory magnetic field called free nuclear precession (FNP). The NV-NMR sensor is then probed using a CASR pulse sequence~\cite{Glenn2018}, which detects the FNP signal and maps it onto a population difference of the NV ensemble electron spin states. The population difference is read out optically by spin state-dependent fluorescence for 1 $\mu$s, followed by optically reinitializing the NV electronic spins for 4 $\mu$s. 

\begin{figure}[htp]
	\centering
	\includegraphics[width=3.4 in]{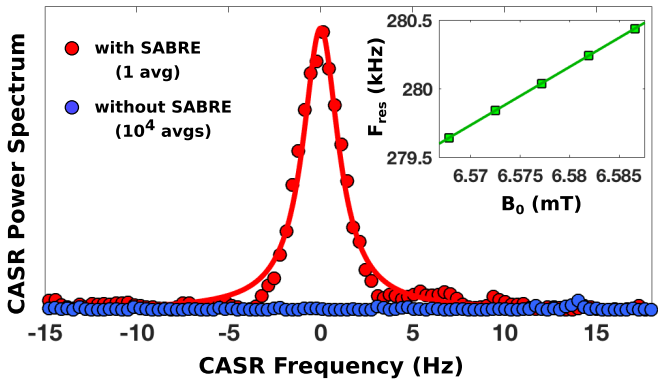} 
	\caption{
		\textbf{SABRE-enhanced NV-NMR spectra of pyridine.} Comparison of measured NV-NMR spectra of 100 mM pyridine sample with (red circles, 1 acquisition) and without (blue circles, 10$^4$ acquisitions) SABRE hyperpolarization using a coherently-averaged synchronized readout (CASR) pulse sequence duration of 2 seconds. The solid red line is a Lorentzian fit to the SABRE-enhanced NV-NMR spectrum, giving a linewidth of 2.3(5) Hz and a signal enhancement of about $2.22(3) \times 10^5$, with a proton number sensitivity of 66.48(15) fmol/$\sqrt{Hz}$ for a signal to noise ratio (SNR) of 3. Inset: CASR resonance frequency F$_{res}$ of hyperpolarized pyridine (green squares) obtained by varying the bias magnetic field B$_0$. A linear fit (green line) of F$_{res}$ versus B$_0$ gives $\gamma_p$ = 42.5355(40) MHz/T, consistent with the proton gyromagnetic ratio. }
	\label{fig:fig2}
\end{figure}

\begin{figure}[hb]
	\centering
	\includegraphics[width=3.4 in]{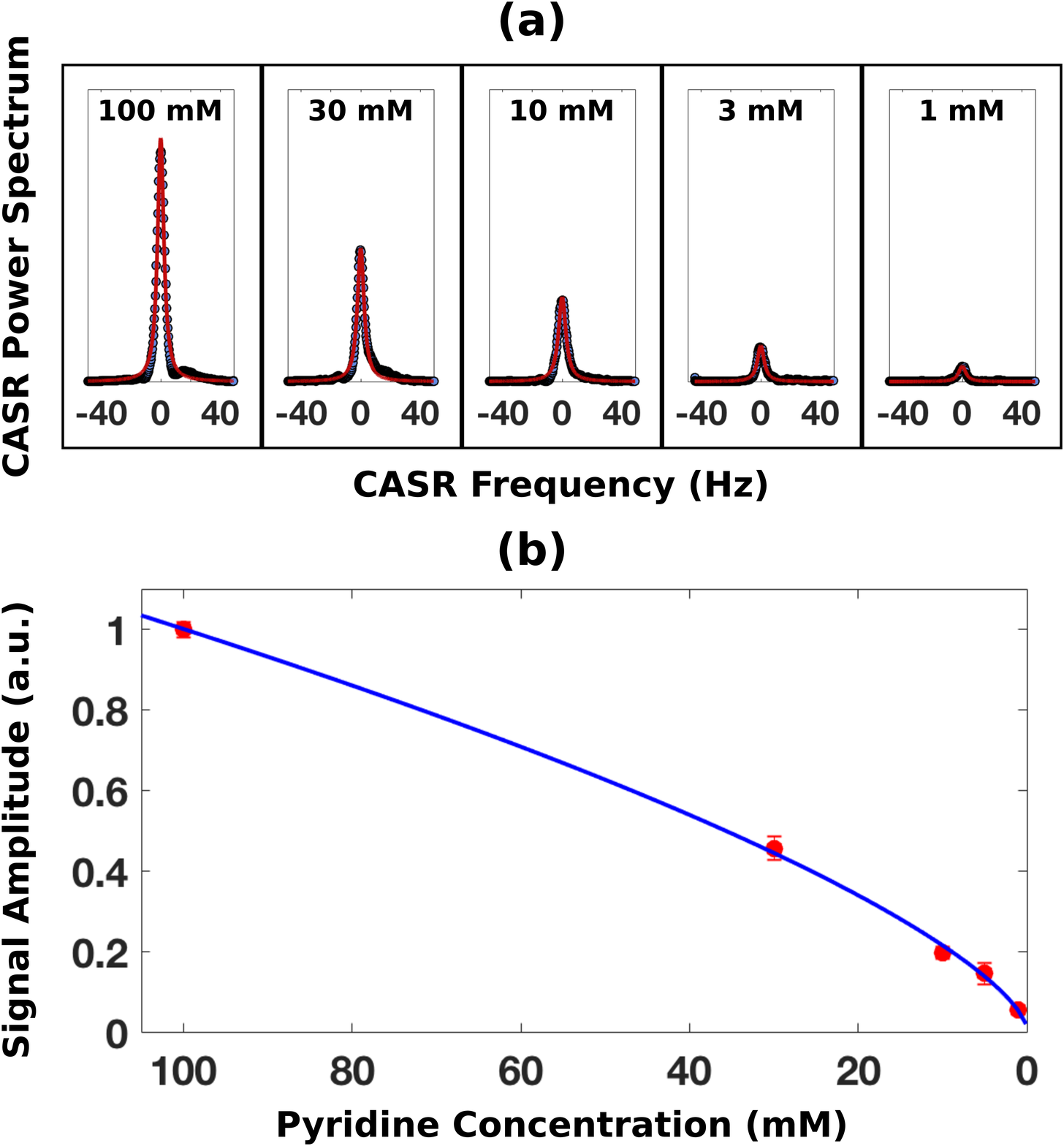} 
	\caption{
		\textbf{SABRE-enhanced NV-NMR measurement for variable pyridine concentrations. (a)} CASR detected NV-NMR spectra (blue circles) for hyperpolarized pyridine samples and its associated Lorentzian fits (solid red line) at concentrations of 100 mM, 30 mM, 10 mM, 5 mM, and 1 mM in methanol. \textbf{(b)} CASR detected NV-NMR signal amplitude (red dots) of hyperpolarized pyridine at various concentrations. The solid blue curve is a power function model of the form $ax^b+c$ with fit parameters a = 0.042, b = 0.6855, c = 0.012.}
	\label{fig:fig3}
\end{figure}

As a first demonstration we detect the SABRE-enhanced NV-NMR spectrum of a sample of pyridine, a weakly-alkaline heterocyclic organic molecule. The sample solution is made with 100 mM concentration of pyridine and 5 mM concentration of catalyst dissolved in methanol. We also apply a calibrated test AC magnetic signal using a coil antenna~\cite{SI}. The observed NV-NMR spectrum with (red dots) and without (blue dots) SABRE hyperpolarization is shown in Fig.~\ref{fig:fig2}. The expected thermally-polarized NV-NMR signal amplitude (without SABRE hyperpolarization) is 32 fT  at a bias magnetic field of 6.6 mT~\cite{SI}. The measured SABRE-enhanced NV-NMR signal has a FWHM line width of 2.3(5) Hz and an amplitude of 7.1(1) nT (by comparing with the amplitude of the test signal~\cite{SI}), which is an enhancement of about $2.22(3) \times 10^5$ in signal amplitude over the expected thermally-polarized signal. The signal to noise ratio (SNR) of this single-shot hyperpolarized NMR signal is 320(3) for a measurement duration of 2 seconds. This result corresponds to a molecule number sensitivity of 13.3(3) fmol/$\sqrt{Hz}$ for pyridine and a proton number sensitivity of 66.48(15) fmol/$\sqrt{Hz}$, which is a two orders of magnitude improvement in proton number sensitivity compared to the Overhauser DNP technique applied to NV-NMR~\cite{Bucher2018}. The sensitivity is defined relative to a signal to noise ratio (SNR) of 3, which is typical in conventional NMR~\cite{Badilita2012}. The pressure, flow rate, and parahydrogen bubbling duration are optimized to achieve this enhancement in sensitivity~\cite{SI}. The NV-NMR signal without hyperpolarization (blue dots in Fig.~\ref{fig:fig2}) is too weak to observe even after $10^4$ averages. We verify the hyperpolarized pyridine NV-NMR signal by measuring the signal resonance frequency F$_{res}$ as a function of applied bias magnetic field B$_0$, yielding a variation of  $\gamma_p$ = 42.5355(40) MHz/T, consistent with the gyromagnetic ratio of the proton.

We next perform SABRE-enhanced NV-NMR spectroscopy (Fig.~\ref{fig:fig3}) by further diluting pyridine in methanol. Samples are prepared at concentrations of 100 mM, 30 mM, 10 mM, 5 mM, and 1 mM of pyridine dissolved in methanol. The ratio of pyridine concentration to the catalyst concentration (20:1) is kept constant~\cite{Colell2017}. A hyperpolarized NV-NMR spectrum is observed even at a concentration of 1 mM (10 femtomoles of pyridine molecules) with a signal-to-noise ratio (SNR) of 50 after averaging for 300 s (Fig.~\ref{fig:fig3}a). The red dots in Fig.~\ref{fig:fig3}b denote the detected NV-NMR signal amplitude at various pyridine concentrations. The error bars represent the standard deviation of the NV-NMR signal measured across three independent trials. A power function model of the form $ax^b+c$ is fit to the experimental data (Fig.~\ref{fig:fig3}b solid blue curve), in excellent agreement with the measurements for fit parameters $a$ = 0.042, $b$ = 0.6855, and $c$ = 0.012. Deviations from a linear dependence are expected since SABRE hyperpolarization of fewer substrate molecules is more efficient than hyperpolarization of more substrate molecules~\cite{Colell2017,Cowley2011,Truong2015}, and thus the relative hyperpolarization decreases with increasing pyridine concentration. The model fit in Fig.~\ref{fig:fig3}b is repeatable and could be used, for a given SABRE NV-NMR system and within a calibrated concentration range, to quantify samples of unknown pyridine concentration.

\begin{figure}[htb]
	\centering
	\includegraphics[width = 3.3 in]{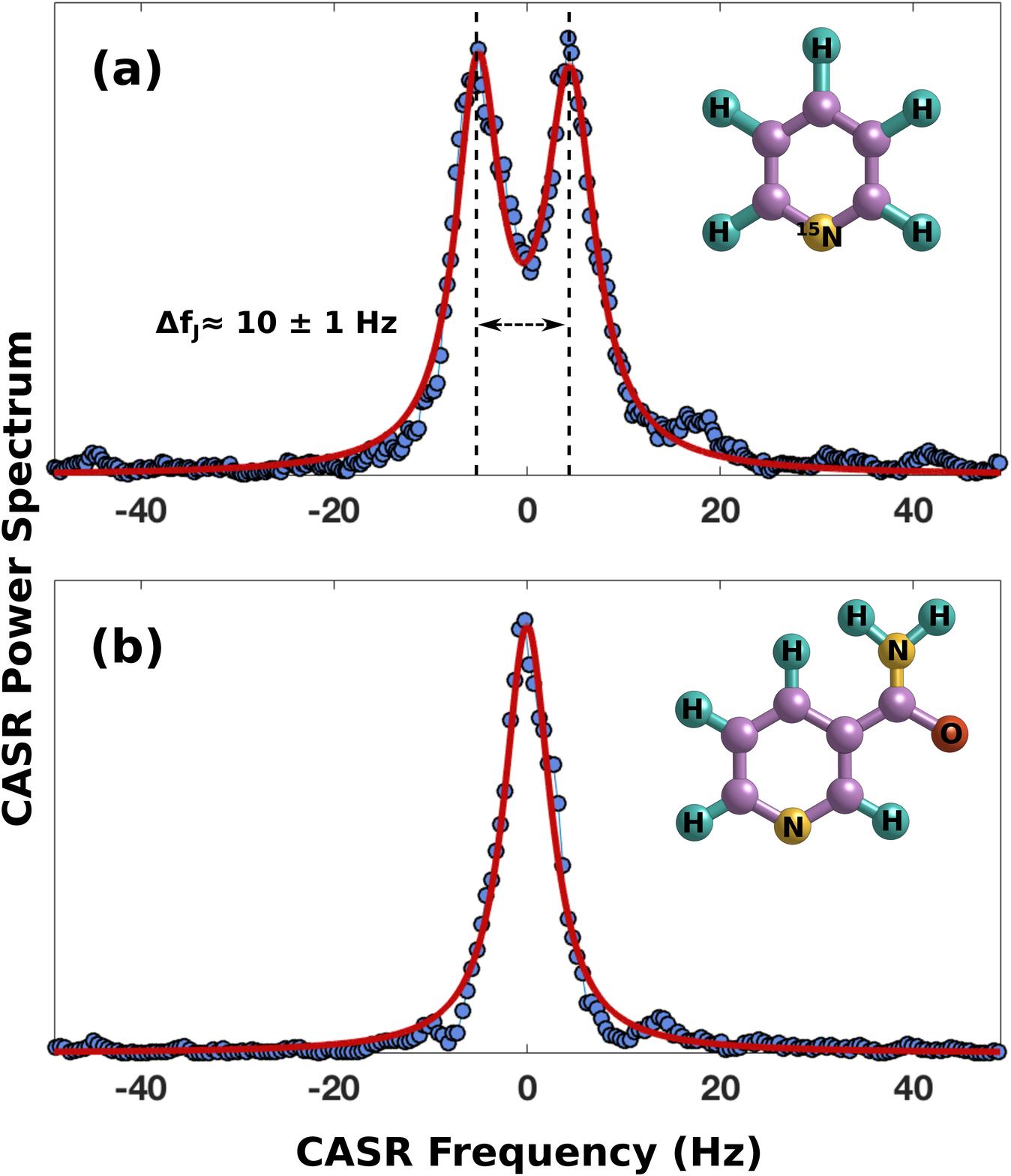}
	\caption{\textbf{SABRE-enhanced molecular NV-NMR spectra. (a)} Single-shot CASR spectrum of hyperpolarized $^{15}$N-labeled pyridine (blue circles) for a sensing duration of 3 seconds at a concentration of 100 mM. The double Lorentzian fit (solid red line) indicates a splitting of $\Delta f_J \approx$ 10(1) Hz (indicated by the vertical dashed lines) due to \textit{J}-coupling between the $^{15}$N nucleus and the protons. Inset: Chemical structure of $^{15}$N-labeled pyridine \textbf{(b)} Single-shot CASR spectrum of hyperpolarized nicotinamide (blue circles) for a sensing duration of 2 seconds at a concentration of 100 mM. The Lorentzian fit (solid red line) gives a spectral linewidth $\approx$ 4(1) Hz. Inset: Chemical structure of nicotinamide.
		\label{fig:fig4}}
\end{figure}

To illustrate the versatility of our technique, we acquire SABRE-enhanced NV-NMR signals from two additional molecules (Fig.~\ref{fig:fig4}). First, we study $^{15}$N-labeled pyridine, which has a \textit{J}-coupling of about 10 Hz between the nuclear spins of the protons and the $^{15}$N~\cite{Theis2015,Truong2015}. The sample is prepared with a 100 mM concentration of $^{15}$N-labeled pyridine and a 5 mM concentration of catalyst dissolved in methanol. The hyperpolarized NV-NMR spectrum has an SNR of 150(5) (Fig.~\ref{fig:fig4}a) for a CASR pulse sequence duration of 3 seconds. This high-resolution spectrum has a linewidth of 3(1) Hz and shows well resolved peaks due to the \textit{J}-coupling~\cite{Lehmkuhl2020}, with a splitting of $\Delta f_J \approx$ 10(1) Hz (Fig.~\ref{fig:fig4}a, solid red line) determined from a double Lorentzian fit. Finally, we measure the SABRE-enhanced NV-NMR spectrum of nicotinamide, a water-soluble form of vitamin B$_3$ (niacin), at 100 mM concentration. The observed NMR spectrum has an SNR of 200(4) (Fig.~\ref{fig:fig4}b) for a CASR pulse sequence duration of 2 seconds. The spectral linewidth from the Lorentzian fit (Fig.~\ref{fig:fig4}b, solid red line) is 4(1) Hz. 

Nicotinamide has important functions in mammalian metabolism and is a metabolic precursor to NAD+/NADH~\cite{Ellinger1949,Canto2012,Sauve883}. With further development, we envision using our technique to observe the conversion of Nicotinamide to NAD+/NADH, which could allow NMR measurement of the redox status in cells. Hyperpolarized NV-NMR may also enable metabolic studies of healthy and diseased cells with dysregulated metabolism on the single-cell level. To increase the chemical specificity necessary for such applications, the SABRE technique can be implemented at a tesla-scale bias magnetic field. For example, in RF-SABRE methods~\cite{Pravdivtsev2015,Theis2018,Ariyasingha2019}, NMR pulse sequences are applied to the catalyst-substrate complex spins, allowing polarization transfer from parahydrogen derived hydrides to substrate molecules at any magnetic field. The current hyperpolarization enhancement can be further increased by implementing the SABRE method on a microfluidic-diamond chip, thereby increasing the contact area between the substrate and the parahydrogen~\cite{Bordonali2019,Smits2019}. A factor of three times improvement in hyperpolarization can also be obtained by utilizing pure parahydrogen instead of the 50$\%$ parahydrogen produced by our home-built system~\cite{SI}.

In summary, we demonstrate about a $10^5$ improvement in NV-NMR proton concentration sensitivity over thermal polarization at 6.6 mT by hyperpolarizing sample proton spins through the technique of signal amplification by reversible exchange (SABRE). This advance augments the growing toolbox of techniques for sensitive, high-resolution NMR spectroscopy in micron-scale samples using NV quantum defects in diamond. Compared to other signal enhancement methods, such as room temperature Overhauser DNP~\cite{Bucher2018} or direct flow-based pre-polarization~\cite{Smits2019}, SABRE provides significantly higher concentration sensitivity while being applicable to a wide range of small molecule analytes~\cite{Colell2017,Rayner2018,Wissam2018}. With planned extension to tesla-scale magnetic fields, SABRE-enhanced NV-NMR may become a high-impact tool for biological applications, such as tracking and monitoring of chemical reactions of metabolites in single cells.

\bibliography{NVNMR_bibfile}

\end{document}


\title{Supplement: Micron-scale SABRE-enhanced NV-NMR Spectroscopy}


\maketitle


\section{Parahydrogen generation}
\label{sec:theory}

\begin{figure*}[htb]
	\centering
	\includegraphics[width=7.0 in]{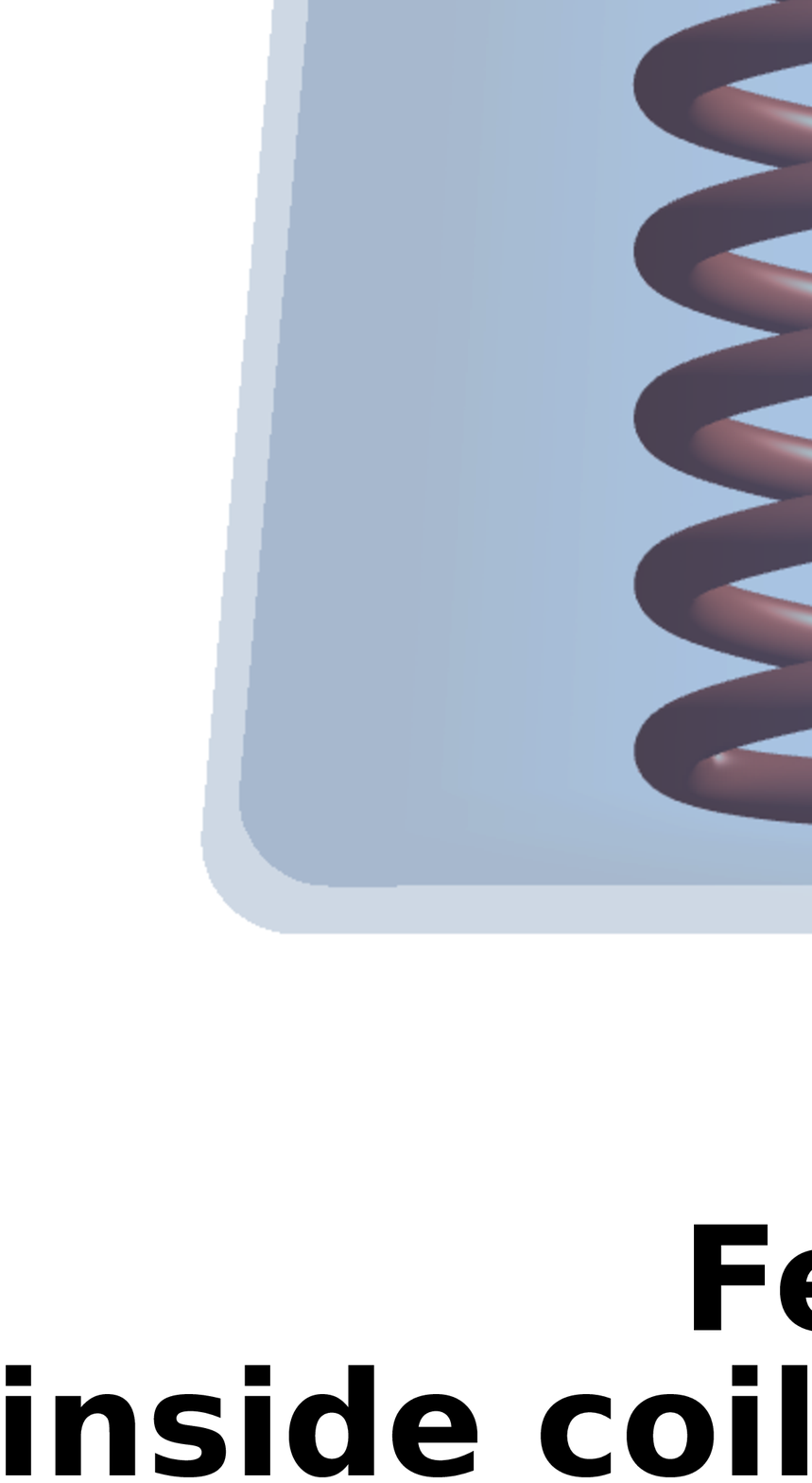} 
	\caption{Schematic of experimental apparatus for parahydrogen production and delivery to NV-NMR set-up ("experimental cuvette"). Parahydrogen is produced by flowing hydrogen gas through an iron oxide catalyst at 77 K. The parahydrogen is then bubbled through the capillary tube located within the experimental cuvette, where it interacts with the SABRE catalyst and hyperpolarization of the small molecule substrate takes place. Optical NV-NMR detection is performed using a diamond chip integrated with the capillary tube in a manner similar to that previously reported~\cite{Glenn2018,Bucher2018}.  Note that other key aspects of the NV-NMR set-up are not shown, including the electromagnet, green laser, optics, and microwave antenna.}
	\label{fig:fig1}
\end{figure*}

We generate parahydrogen using a home-built system, as illustrated in Fig.~\ref{fig:fig1}. Room temperature hydrogen gas, consisting of of 25$\%$ para and 75$\%$ ortho hydrogen, fills a copper tube filled with an iron oxide hydroxide (FeO(OH)) catalyst bed. The copper tube is then submerged in liquid nitrogen at approximately 77K. The iron oxide hydroxide catalyst induces rapid thermalization by allowing interconversion of orthohydrogen to parahydrogen on the catalyst surface. Following enrichment at low temperature for one minute, the gas can be handled at room temperature for several hours (up to days) without loss of the enriched parahydrogen fraction. The parahydrogen gas is then bubbled at a rate of $\sim$ 75 sccm through the SABRE solution in an optical cuvette, and the small molecule substrate is thereby hyperpolarized (details in Section~\ref{sec:SABREprocess}). The parahydrogen gas is maintained at a room temperature pressure of 30 PSI before the flow controller, for all the experiments described in the main text.

\section{SABRE hyperpolarization process}
\label{sec:SABREprocess}

\subsection{\textbf{Chemical Activation}}
\label{sec:Chemicalactivation}

\begin{figure*}[htb]
	\centering
	\includegraphics[width=6.5 in]{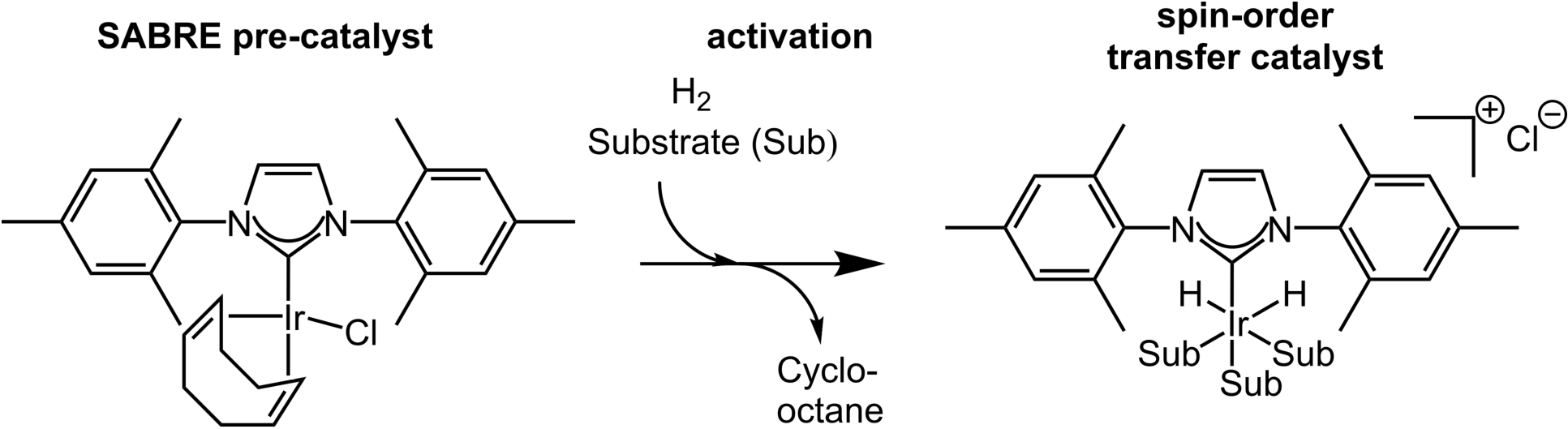} 
	\caption{Illustration of the chemical activation of the SABRE-catalyst precursor to become the SABRE active spin-order transfer catalyst. }
	\label{fig:fig2}
\end{figure*}

The Iridium based catalyst used in SABRE hyperpolarization is ([IrCl(COD)(IMes)] (COD = 1,5-cyclooctadiene; IMes = 1,3-bis (2,4,6-trimethylphenyl)-imidazol-2-ylidene)). As illustrated in Fig.~\ref{fig:fig2}, the SABRE catalyst precursor is chemically activated by supplying hydrogen gas and substrate (e.g., pyridine or nicotinamide). Hydrogen undergoes oxidative addition onto the iridium and the COD (cyclooctadiene) in the catalyst precursor is hydrogenated to cyclooctane.  Hence the COD will no longer interact with the catalyst. Instead, the substrate coordinates with the Iridium to form the active spin-order transfer catalyst. Although enriched parahydrogen gas is used to activate the catalyst for experimental convenience, the spin state of the hydrogen gas is irrelevant during the activation process.

\subsection{\textbf{Reversible Exchange}}
\label{sec:Reversibleexchange}

\begin{figure*}[htb]
	\centering
	\includegraphics[width=6.5 in]{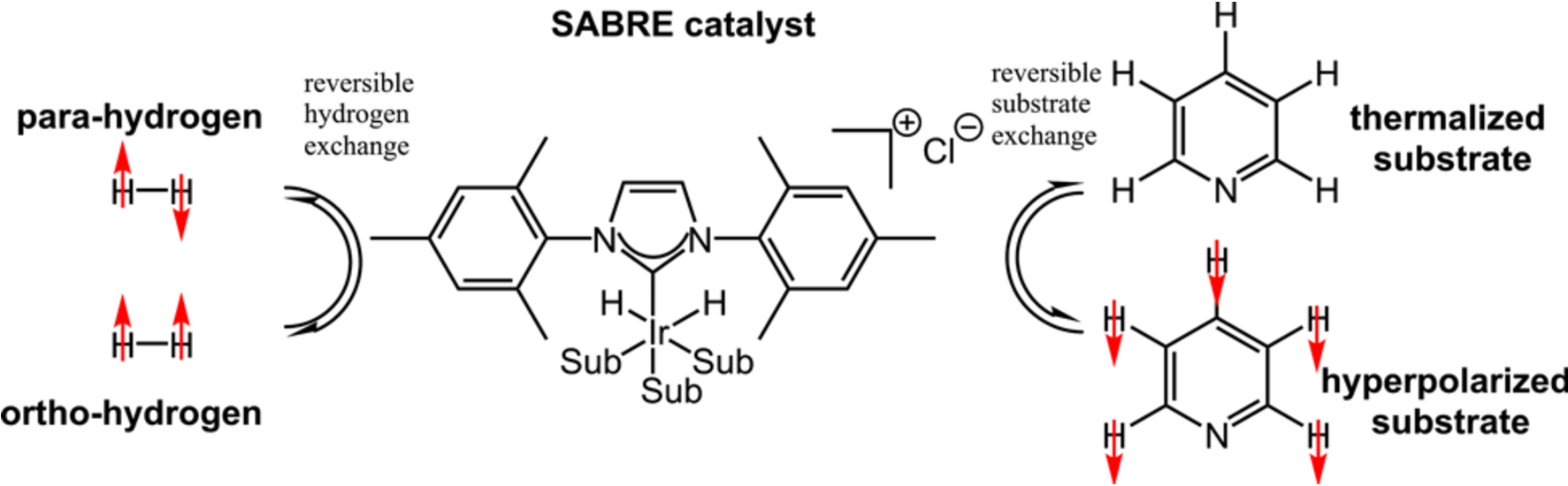} 
	\caption{Illustration of the reversible hydrogen exchange that leads to hyperpolarization buildup on the free molecular substrate in solution. Left: free (unbound) hydrogen in solution. Center: SABRE active spin-order transfer catalyst. Right: free (unbound) substrate in solution. Steady-state excess of parahydrogen in sample solution, maintained by periodic bubbling, leads to steady-state excess of hyperpolarized free substrate.}
	\label{fig:fig3}
\end{figure*}

After chemical activation of the catalyst (Section~\ref{sec:Chemicalactivation}), parahydrogen and the small molecule substrate undergo reversible exchange with the catalyst (Fig.~\ref{fig:fig3}). In the sample solution there are free (unbound) hydrogen, free (unbound) substrate, and the SABRE active spin-order transfer catalyst. During the lifetime of the catalyst (on
the order of ms), spin-order can flow from parahydrogen to the substrate (pyridine in Fig.~\ref{fig:fig3}). Spin-order transfer is most efficient in an ambient magnetic field of 6.6 mT (details in Section~\ref{sec:Polarizationtransfer}). The reversible exchange process and spin-order transfer act together to continually hyperpolarize the free molecular substrate, as long as a steady-state excess of parahydrogen is maintained in the sample solution by periodic bubbling. 

\subsection{\textbf{Spin-order transfer at level-anti-crossing }}
\label{sec:Polarizationtransfer}

\begin{figure*}[htb]
	\centering
	\includegraphics[width=4.7 in]{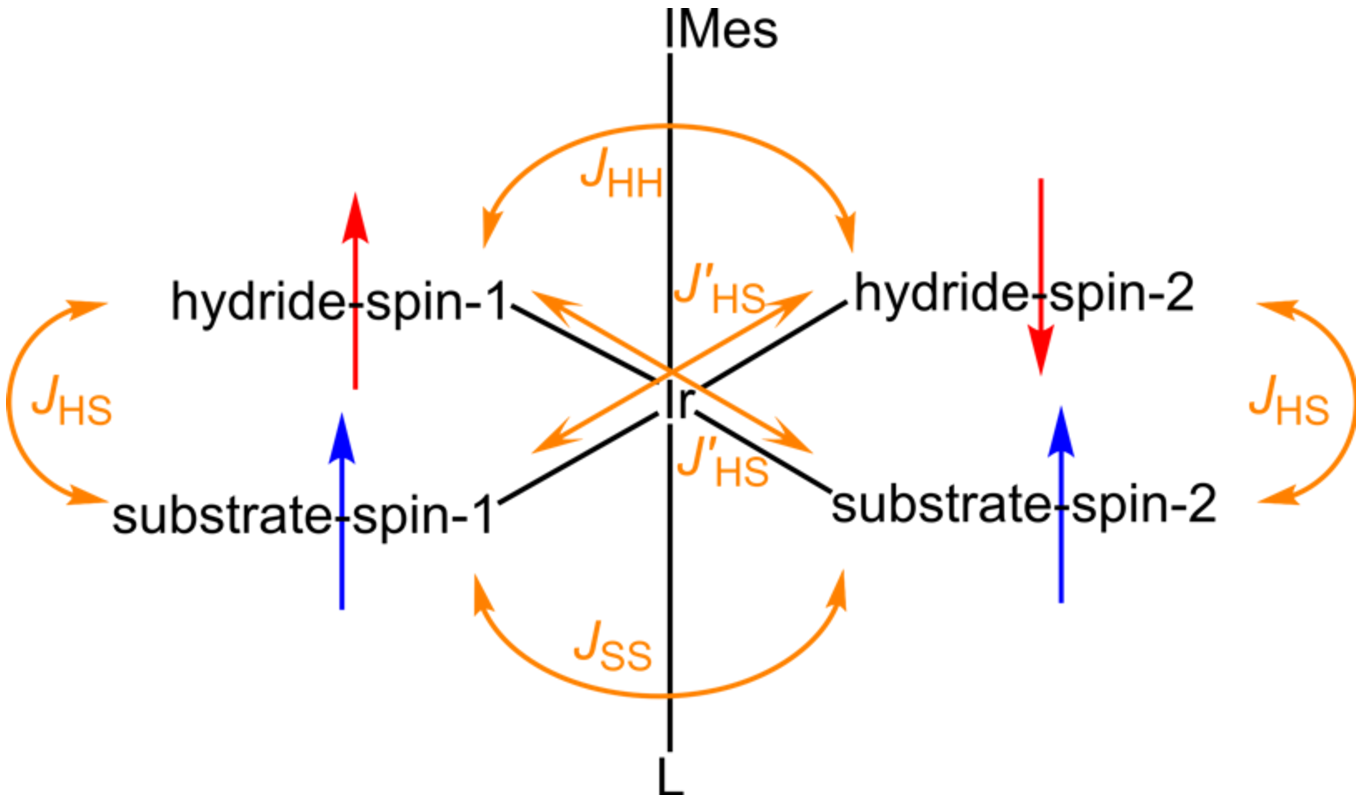} 
	\caption{Illustration of a simplified  spin system that promotes spin-order transfer on the SABRE catalyst. The \textit{J}-coupling values are approximately as follows: \textit{J}$_{HH}\approx$ -7 Hz, \textit{J}$^\prime_{HS}\approx$ 1.5 Hz, \textit{J}$_{HS}\approx$ 0 Hz, \textit{J}$_{HS}\approx$ 0 Hz.}
	\label{fig:fig4}
\end{figure*}

As a SABRE example, two nuclear spins from the hydrides and five nuclear spins from each of two pyridine substrate molecules participate in the spin-order transfer process. Understanding this system of twelve nuclear spins is difficult~\cite{Theis2015}. A simplified model, with only four spins, nonetheless allows us to analytically predict the resonant magnetic field for the spin-order transfer process with reasonable accuracy (see Fig.~\ref{fig:fig4}). To build this model, we begin by constructing the most appropriate eigenstates for the spin system at hand. We choose to combine the singlet ($S^H$) - triplet ($T_+^H,T_0^H,T_-^H$) system of the hydride spins with the singlet ($S^S$) - triplet ($T_+^S,T_0^S,T_-^S$) system of the substrate spins. As a result, we obtain 16 eigenstates (2$^n$ with $n$=4 spins), which we name with an uncoupled nomenclature as follows:
\begin{eqnarray*}
    &S^HS^S,S^HT_+^S,S^HT_0^S,S^HT_-^S,\\
    &T_+^HS^S,T_+^HT_+^S,T_+^HT_0^S,T_+^HT_-^S,\\
    &T_0^HS^S,T_0^HT_+^S,T_0^HT_0^S,T_0^HT_-^S,\\
    &T_-^HS^S,T_-^HT_+^S,T_-^HT_0^S,T_-^HT_-^S
\end{eqnarray*}

The Hamiltonian of the spin system can be expressed in this basis as described in the supplement of~\cite{Theis2015}. The eigenstates responsible for the spin-order transfer at level anti crossing are $S^HS^S$ and $T^H_+T^S_-$. These states are chosen because they illustrate a connection between the parahydrogen derived singlet state on the hydrides ($S^H$) with a magnetization state on the substrate ($T^S_-$). The states are connected in a 2x2 block of the Hamiltonian as follows:
\renewcommand{\arraystretch}{2.5}
\[
\begin{blockarray}{r c c}
& \ket{S^HS^S} & \ket{T^H_+T^S_-}\\
	\begin{block}{r[>{\quad}c c<{\quad}]}
		\ket{S^HS^S} {\quad} & -\textit{J}_{HH}+\textit{J}_{SS} & \dfrac{\Delta\textit{J}_{HS}}{2}
		\\
		\ket{T^H_+T^S_-} {\quad} &\dfrac{\Delta\textit{J}_{HS}}{2} & \nu_H-\nu_S-\Sigma\textit{J}_{HS}
		\\
	\end{block}
\end{blockarray}
\]

where \textit{J}$_{HH}$ is the \textit{J}-coupling between the hydrides, \textit{J}$_{SS}$ is the \textit{J}-coupling between the substrate spins, $\nu_S$ is the Larmor frequency of the substrate spins, $\nu_H$ is the Larmor frequency of the hydride spins, $\Delta\textit{J}_{HS}=\textit{J}_{HS}-\textit{J}^\prime_{HS}$, and $\Sigma\textit{J}_{HS}=\textit{J}_{HS}+\textit{J}^\prime_{HS}$. When the diagonal matrix elements are equal to one another, level-anti-crossing is established and $\Delta\textit{J}_{HS}$ can most efficiently transfer spin-order from the hydride singlet state $S^H$ to the substrate $T^S_-$ state. 

With this Hamiltonian we can now, from first principles, determine the magnetic field at which spin-order flow is most efficient, as follows. Equating the diagonal elements gives,
\begin{equation}
-\textit{J}_{HH}+\textit{J}_{SS} =  \nu_H-\nu_S-\Sigma\textit{J}_{HS}.
\label{eqn:equ1}
\end{equation}
Solving for the frequency difference $\nu_H-\nu_S$ in Eq.~\ref{eqn:equ1} gives,
\begin{equation}
\nu_H-\nu_S=  \Sigma\textit{J}_{HS}-\textit{J}_{HH}+\textit{J}_{SS}. 
\label{eqn:equ2}
\end{equation}
The frequency difference $\nu_H-\nu_S$ also depends on the magnetic field B as
\begin{equation}
\nu_H-\nu_S=  \gamma B(\delta_H-\delta_S)
\label{eqn:equ3}
\end{equation}
where $ \gamma$ is the gyromagnetic ratio of hydrogen, $\delta_H$ and $\delta_S$ are the chemical shift of the hydrides and the substrate, respectively.
Solving for the magnetic field B in Eq.~\ref{eqn:equ3} yields,
\begin{equation}
B= \dfrac{ \nu_H-\nu_S}{\gamma (\delta_H-\delta_S)}.
\label{eqn:equ4}
\end{equation}
Comparing Eq.~\ref{eqn:equ2} and Eq.~\ref{eqn:equ4}, we obtain
\begin{equation}
B= \dfrac{\Sigma\textit{J}_{HS}-\textit{J}_{HH}+\textit{J}_{SS}}{\gamma (\Delta\delta)}.
\label{eqn:equ5}
\end{equation}
where $\Delta\delta= \delta_H-\delta_S$.

For our SABRE system, $\Sigma\textit{J}_{HS}$ = 1.5 Hz; $\textit{J}_{HH}$ = 7 Hz; $\textit{J}_{SS}$ = 0; $\Delta\delta$ = 30 ppm. Substituting theses values in Eq.~\ref{eqn:equ5} gives 6.65 mT, the ambient magnetic field at which hyperpolarization is most efficient. The experiments shown in the main text are implemented at a magnetic field of 6.576 mT.

\section{Experimental Methods}
\label{sec:methods}

Details about the NV ensemble sensor, magnetic bias field stabilization, NMR drive coils, synchronized readout protocol, and data analysis are described in the Methods section of~\cite{Glenn2018}. The 200 mW optical beam ($\lambda$ = 532 nm), generated by a solid-state laser (Coherent Verdi G7), is focused down to a spot size of about 15 $\mu$m and pulsed using an acousto-optic modulator (AOM) (IntraAction ASM802B47). The duration of each optical pulse is 5 $\mu$s. The NV spin-state-dependent fluorescence is read out for 1 $\mu$s followed by optically reinitializing the NV electronic spins for 4 $\mu$s. 

The experimental sequence for the SABRE hyperpolarization with CASR NV-NMR detection is illustrated in Fig. 1d of the main text. Shortly after preparing the sample, parahydrogen is dispersed into the sample solution for 20 minutes to activate the SABRE catalyst. This step is performed only once. Once the catalyst is activated, parahydrogen is bubbled into the sample for 30 seconds, before every CASR NV-NMR measurement, to hyperpolarize the proton nuclear spins in the sample. After a 1 second wait time for the bubbles to settle, NV-NMR detection of the hyperpolarized sample is performed using the CASR measurement protocol~\cite{Glenn2018}. 

\begin{figure*}[htb]
	\centering
	\includegraphics[width=5.9 in]{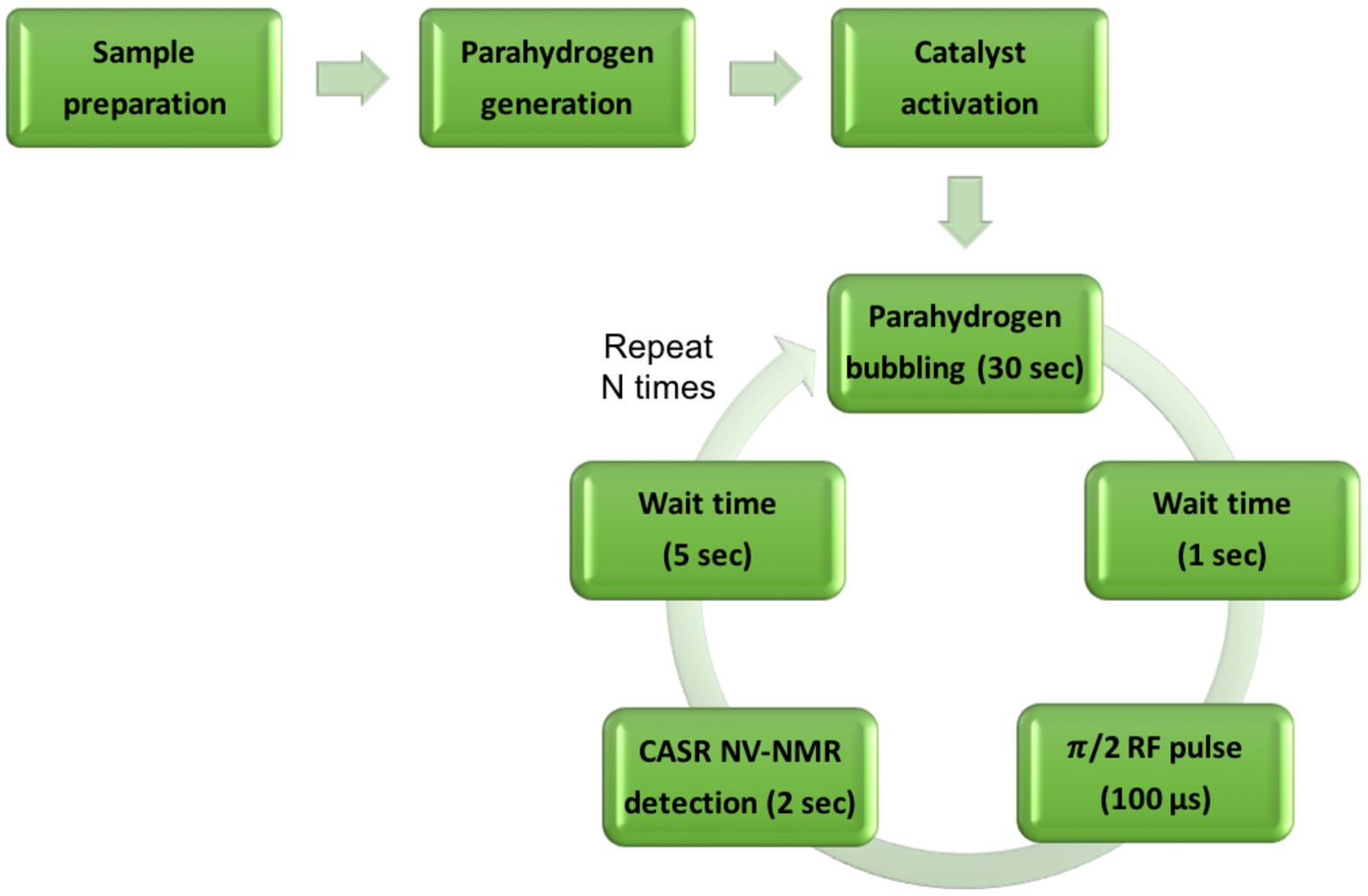} 
	\caption{Experimental protocol. Experimental process like sample preparation, parahydrogen generation, and catalyst activation are performed only once. NMR measurement, including parahydrogen bubbling and CASR NV-NMR measurement can be repeated N times, where N is the number of averages used to improve SNR.}
	\label{fig:fig5}
\end{figure*}

A CASR NV-NMR measurement is summarized here.  First, a $\pi$/2 RF pulse ($\sim$ 280 kHz) with a duration of 100 $\mu$s is applied by the NMR drive coils, resonant with the proton spins of the sample, to induce free nuclear precession (FNP). The resulting FNP signal from the sample is then measured with CASR NV-NMR detection. The CASR pulse sequence, applied to the NV ensemble spins, is programmed on an arbitrary waveform generator (Tektronix AWG 7122C) and triggered by a pulse generator (Spincore PulseBlaster ESR-PRO 500 MHz). The CASR pulse sequence consists of interspersed blocks of identical XY8-1 subsequences, each followed by optical readout of the ensemble NV spins, with a total duration of about 25 $\mu$s per XY8-1 subsequence and readout. The $\pi$ and $\pi$/2 pulse durations used in the XY8-1 subsequence are 60 ns and 30 ns, respectively. To remove laser and MW noise, the last $\pi$/2 pulse on every alternate XY8-1 subsequence is phase-shifted by $\pi$ and the NV fluorescence measurements from the alternating XY8-1 subsequences are subtracted from each other. Hence two consecutive XY8-1 subsequences ($\sim$ 50 $\mu$s) yield one data point of CASR NV-NMR detection, with a total measurement typically consisting of 40,000 points ($\sim$ 2 seconds total measurement time).  After a CASR NV-NMR measurement, there is a 5 seconds wait time ($\sim$ T$_1$ of the hyperpolarized proton spins in the sample) before starting the next experimental sequence (i.e., 30 seconds of parahydrogen bubbling, 1 second wait time, and CASR NV-NMR measurement for 2 seconds). The experiment protocol is summarized in Table~\ref{tab:Exprotocol} and Fig.~\ref{fig:fig5}.

\begin{table}[htb]
\caption{Experiment Protocol}  
\centering  
\setlength{\tabcolsep}{8pt}
\begin{tabular}{c c c c c}  
\hline\hline                       
 & Steps & Location &Duration & Temperature \\ [0.5ex]   
\hline\hline              
&Sample preparation& - & - & Room temperature \\[0ex]
\raisebox{0.1ex}{Preparation} & Parahydrogen generation & \parbox[m]{4cm}{Inside coiled copper tube \\filled with [FeO(OH)]}& $\sim$ 1 minute &$\sim$ 77 K \\[0ex]
& Catalyst activation & \parbox[m]{4cm}{Inside the experimental\\cuvette} & 20 minutes & Room temperature \\[1ex]
\hline              
&Parahydrogen bubbling & & 30 seconds & \\[-1ex]
&Wait time & & 1 second & \\[-1ex]
\raisebox{3ex}{NMR Measurement} & CASR NV-NMR detection & \raisebox{3ex} {\parbox[m]{4cm}{Inside the experimental\\cuvette}} & 2 seconds & \raisebox{3ex}{Room temperature} \\[-1ex]
&Wait time & & 5 seconds & \\[1ex]
\hline\hline                         
\end{tabular}
\label{tab:Exprotocol}
\end{table}

\section{Determination of SABRE NV-NMR signal enhancement}
\label{sec:enhancement}

Determination of the NMR signal enhancement using SABRE hyperpolarization of the NV ensemble magnetometer is carried out in two steps: (i) measuring the NV-NMR signal amplitude with the SABRE technique; and (ii) calculating the expected NV-NMR signal amplitude without SABRE.

\begin{figure*}[htb]
	\centering
	\includegraphics[width=6.5 in]{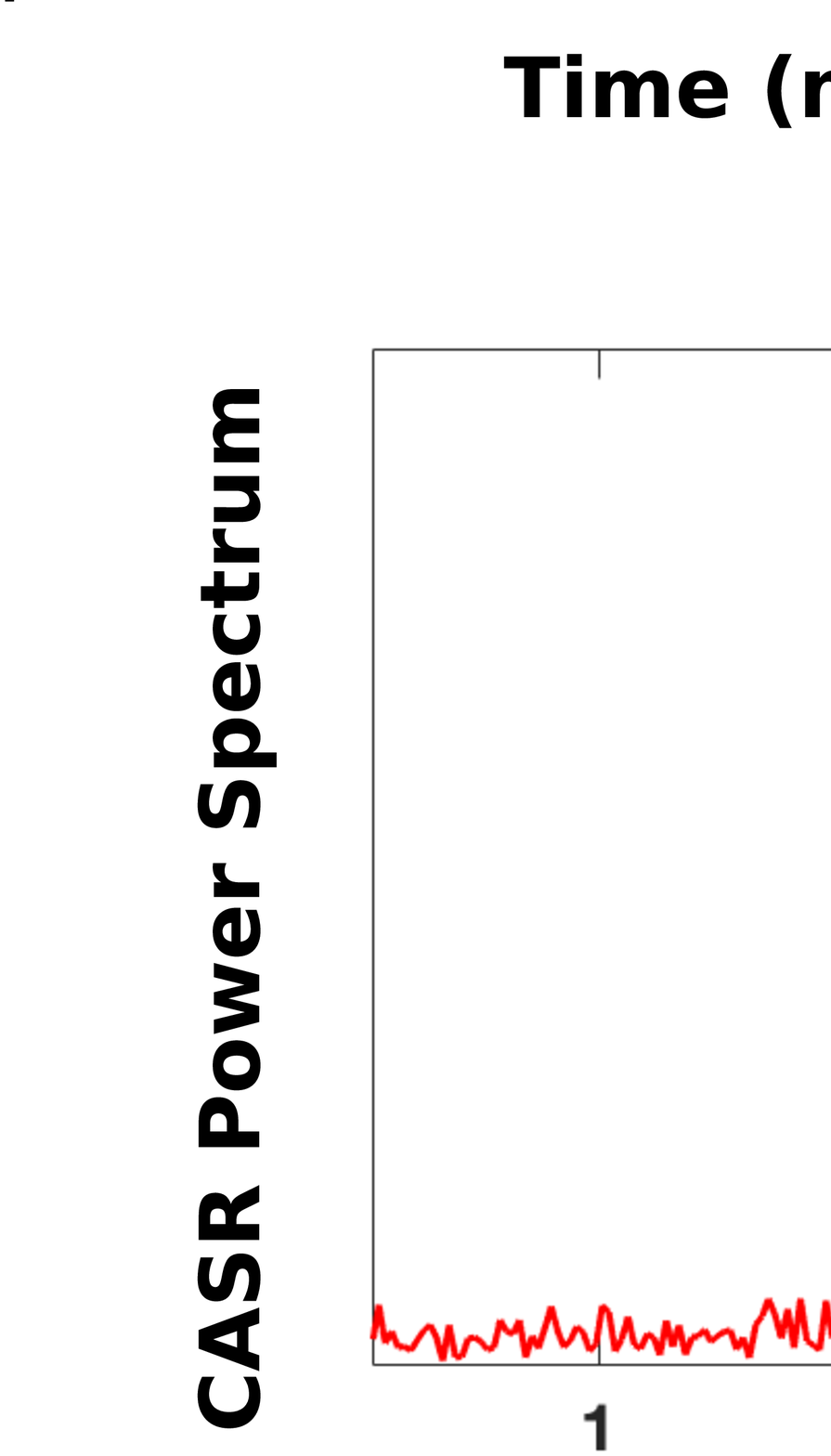} 
	\caption{\textbf{(a)} Coherently averaged synchronized readout (CASR) NV measurements of the AC magnetic test signals from a nearby coil at $f_{coil}$ = 280 kHz. The coil drive voltage V$_c$ for the AC signal source of the coil was varied from V$_c$ = 0 (purple trace) to V$_c$ = 0.6 V (magenta trace). \textbf{(b)} CASR signal data (blue points) as a function of coil voltage V$_c$ at time t = 0.302 ms, obtained along the dashed line in Fig.~\ref{fig:fig6}a. The blue line is a sinusoidal fit to the data, which gives the magnetic test signal calibration of 5.7(2) $\mu$T/V. \textbf{(c)} CASR detected NV-NMR spectra (peak A) of pyridine and test signal (peak B).}
	\label{fig:fig6}
\end{figure*}

\subsection{\textbf{Measurement of the NV-NMR signal amplitude with SABRE}}
\label{sec:NMRwithSABRE}

We measure the magnitude of the NV-NMR signal amplitude by comparing (a) the CASR fluorescence signal from the SABRE-hyperpolarized sample, as discussed in the main text, to (b) the CASR signal from an AC magnetic field generated from a test coil. The AC magnetic field generated by the test coil is first calibrated, as summarized here and in~\cite{Glenn2018}. The test AC field is of the form 
\begin{equation}
b(t) = b_{ac} \sin (2\pi f_{coil}+\phi),
\end{equation}
where $b_{ac}$ is the magnetic field amplitude that needs to be calibrated, $f_{coil}$ = 280 kHz is the drive frequency, which is within a few kHz of the frequency $f_0$ used in CASR NV-NMR measurements, and $\phi$ is the phase with respect to the first CASR magnetometry sequence. The field amplitude $b_{ac}$ is linearly proportional to the voltage V$_c$ supplied to the test coil antenna. The CASR fluorescence signal from the AC test field (Fig.~\ref{fig:fig6}a) is acquired by varying the test coil drive voltage V$_c$ from 0 V (purple trace) to 0.6 V (magenta trace), thereby varying the test field amplitude $b_{ac}$. The CASR fluorescence signal amplitude at t = 0.302 s in Fig.~\ref{fig:fig6}a (along the dotted line) is plotted as a function of voltage V$_c$  in Fig.~\ref{fig:fig6}b (blue circles). The sinusoidal fit to the data (blue line in Fig.~\ref{fig:fig6}b) results in the test signal calibration of 5.7(2) $\mu$T/V. For the experiments presented in the main text, we applied a test signal with voltage V$_c$ = 1 mV, which corresponds to the test signal amplitude of 5.7(2) nT (Fig.~\ref{fig:fig6}c peak B). Comparing the measured CASR signal from pyridine (Fig.~\ref{fig:fig6}c peak A) with this test signal yields an NMR signal amplitude of 7.1(1) nT.

\subsection{\textbf{Calculation of the NV-NMR signal amplitude without SABRE}}
\label{sec:NMRwithoutSABRE}

The calculated thermal NV-NMR signal amplitude (i.e., without SABRE hyperpolarization) of protons in the pure water at 88 mT and room temperature corresponds to 95 pT for our experimental set-up~\cite{Glenn2018}. The proton molar concentration of pure water is 111 M. Hence for a 100 mM sample concentration, the expected NV-NMR signal amplitude is 85.6 fT. Our experiments are performed in a static magnetic field of 6.576 mT. Since the thermal spin polarization is linearly proportional to the magnetic field in which the experiment is performed, the expected NV-NMR signal amplitude at 6.576 mT corresponds to 6.4 fT. Since pyridine has 5 protons, the expected NV-NMR signal amplitude for thermally-polarized pyridine at 100 mM concentration is $\approx$ 32 fT.
 
\subsection{\textbf{SABRE NV-NMR signal enhancement}}
\label{sec:NMRenhancement}

We determine a $2.22(3) \times 10^5$ enhancement in NV-NMR signal amplitude due to SABRE hyperpolarization, for a pyridine sample of 100 mM concentration, by taking the ratio of the measured NV-NMR signal amplitude of 100 mM pyridine with SABRE (from Section~\ref{sec:NMRwithSABRE}) to the calculated NV-NMR signal amplitude of 100 mM pyridine without SABRE (from Section~\ref{sec:NMRwithoutSABRE}).

\bibliography{NVNMR_bibfile}